\documentclass[format=sigchi, review=false, screen=true,nonacm=true,timestamp=false]{acmart}

\usepackage{booktabs} % For formal tables

\usepackage[ruled]{algorithm2e} % For algorithms

\SetAlFnt{\small}
\SetAlCapFnt{\small}
\SetAlCapNameFnt{\small}
\SetAlCapHSkip{0pt}
\IncMargin{-\parindent}

\begin{document}
% Title portion. Note the short title for running heads
\title[Machine Learning and Fuzzing Survey]{A Review of Machine Learning Applications in Fuzzing}

\author{Gary J Saavedra}
\author{Kathryn N Rodhouse}
\author{Daniel M Dunlavy}
\author{Philip W Kegelmeyer}

\affiliation{%
  \institution{Sandia National Laboratories}
  \streetaddress{1515 Eubank Blvd SE}
  \city{Albuquerque}
  \state{NM}
  \postcode{87123}
  \country{USA}
}
\email{{gjsaave,knrodho,dmdunla,wpk}@sandia.gov}

\begin{abstract}
Fuzzing has played an important role in improving software development and testing over the course of several decades.  Recent research in fuzzing has focused on applications of machine learning (ML), offering useful tools to overcome challenges in the fuzzing process.  This review surveys the current research in applying ML to fuzzing.  Specifically, this review discusses successful applications of ML to fuzzing, briefly explores challenges encountered, and motivates future research to address fuzzing bottlenecks. 
\end{abstract}


\maketitle

%\renewcommand{\shortauthors}{G. Saavedra et al.}

\section{Introduction}

Fuzzing is a technique in which a large number of generated inputs, both valid and invalid, are fed into a program to search for flaws and vulnerabilities. 
Fuzzers, or automated tools to perform fuzzing, have played an important role in quality assurance, system administration, and vulnerability assessment over the last three decades \cite{takanen_08,duran,duran_ntafos,barton_miller}. 
Modern fuzzers now incorporate techniques from other disciplines; in this survey, we explore how some modern fuzzers incorporate different types of machine learning (ML).
We specifically focus on fuzzers used for vulnerability assessment due to their widespread use. 

Machine learning is a method for training computer models to perform some operation without being explicitly programmed. 
ML techniques are applied across a wide range of problems, from image processing to sequence modeling \cite{hinton, goodfellow, sutskever}.
In this survey, we focus on three main types of ML, each of which is suited for a different type of task. 
\textit{Supervised learning} is used to train a model
to identify the \textit{class} label  
of a given data point, such as whether an image contains a specific object. 
This type of ML requires data sets where each data point is given an explicit label. 
\textit{Unsupervised learning} is used to
train a model to
find patterns or similarities between data points rather than labeling classes. 
This type of ML is used when the data does not have explicit labels.
\textit{Reinforcement learning} is used to train a model, often referred to as an \textit{agent}, to take an optimal set of actions in an environment.
This type of ML rewards the agent for each action it takes in the environment.  Similar to supervised learning, the rewards act as a label for the agent and provide an indication of the optimal actions to take.  Thus, the agent can be trained to take a set of actions which lead to the highest reward.
Each of these three types can also take on a particular form of ML called \textit{deep learning}.  Deep learning refers to a type of hierarchical learning that can be used to learn the underlying features and structure of a set of data points \cite{goodfellow}.   

In this survey, we explore how ML has been applied to address core research questions in fuzzers used for vulnerability assessment of programs. 
This survey is representative rather than exhaustive; we do not attempt to discuss all applications of ML in fuzzers. 
In Section \ref{fuzzing}, we give a brief overview of the fuzzing process. 
In Section \ref{ML}, we explore previous and ongoing research applying ML techniques to fuzzing.  We also discuss difficulties associated with applying ML to fuzzing.
We conclude with a summary of how ML has been applied to fuzzing and how ML might be applied to fuzzing in the future.

\section{Overview of fuzzing}
\label{fuzzing}

In ``A Review of Fuzzing Tools and Methods'' \cite{fell}, Fell explores the use of fuzzers for vulnerability assessment.
We recommend Fell's treatment of fuzzing for the interested reader as he describes the process in depth.
Here, we provide only a brief overview of fuzzing to frame our discussion of where and how machine learning (ML) techniques have been applied, which will be covered in Section \ref{ML}.

Again, fuzzing is a technique in which a large number of generated inputs, both valid and invalid, are fed into a program to search for flaws and vulnerabilities.
Fuzzers automate much of this process, taking in initial program knowledge and reporting out any interesting program states discovered \cite{oehlert_sap_05}. 
Often, a human user both provides initial program knowledge and analyzes these output program states.
Traditionally, ``interesting program states'' were program crashes that identified flaws and vulnerabilities in the program, but more complex program monitoring techniques now allow for identification of other types of interesting states.

The goal of a fuzzer is to create inputs that cause the program to execute program paths, discovering those that lead to interesting program states. 
Thus, fuzzers are frequently measured by the diversity of program paths explored, termed \textit{coverage}. 

In this section, we introduce types of fuzzers and break down the fuzzing process into stages.
We walk through each stage, showing how the different types of fuzzers approach some stages differently. 
We introduce the types of tasks that exist in each stage, framing the most compelling challenges that are currently not fully solved at each stage.
Finally, we touch on ways to compare fuzzers.

\subsection{Types of Fuzzers}
\label{sec:fuzzer_types}

A fuzzer that generates completely random input and feeds it to a program is a \textit{na\"ive fuzzer}.  
While na\"ive fuzzers are fairly easy to implement, they are unlikely to reach interesting program states in a timely manner. 
Three primary types of modern fuzzers improve on na\"ive fuzzers:
\textit{mutation-based}, \textit{generation-based}, and \textit{evolutionary}.

\textbf{Mutation-based fuzzers} blindly \textit{mutate} or manipulate provided input to feed to the program.
Generally mutation-based fuzzers are not aware of the expected input format or specifications, and they cannot select mutations wisely.
Peach is a 
fuzzer that can perform mutation-based as well as generation-based fuzzing \cite{Peach}.

\textbf{Generation-based fuzzers} take information about the expected input format or protocol through specifications.
Generation-based fuzzers \textit{generate} or craft inputs based on these specifications.
Generation-based fuzzers include Peach and Sulley, a Python fuzzing framework that can generate inputs for file transfer protocols, network protocols, and file formats \cite{Sulley}.

\textbf{Evolutionary fuzzers}, the newest type of fuzzers, build on mutation-based fuzzers by selecting some inputs over others for mutation.
Specifically, evolutionary fuzzers aim to evaluate what each input causes the program to do and change how they proceed based on that evaluation.
Practically, current state-of-the-art evolutionary fuzzers rank inputs using a fitness function (often coverage) and select the best-ranked inputs to mutate.
Example evolutionary fuzzers include honggfuzz \cite{honggfuzz}, AFL \cite{AFL}, and libFuzzer \cite{libfuzzer}.  

For the interested reader, Vimpari's 2015 thesis evaluates the utility of some free fuzzers \cite{vimpari}. 
However, we move on to explore the fuzzing process.

\begin{figure*} 
  \includegraphics[width=\textwidth]{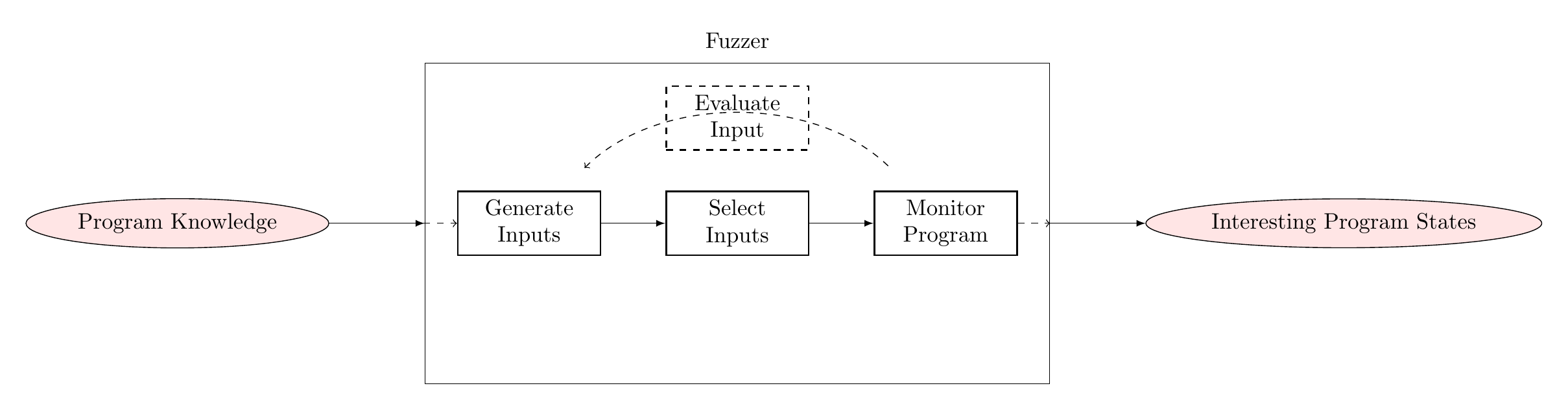}
  \caption{Fuzzing Process}
\end{figure*}

\label{fuzzing_workflow}
\subsection{Fuzzing Process}
Figure 1 breaks a general fuzzing process into stages.
Every fuzzer takes in \textit{program knowledge}, uses that knowledge to \textit{generate inputs}, \textit{select inputs} to feed to the program being fuzzed, \textit{monitor program} actions in response to each input, and identify and export any \textit{interesting program states} for the user to analyze.
The fuzzer automatically repeats the inner stages, and it can use results from monitoring the program to \textit{evaluate input} and inform the next input generation stage.
While this process describes fuzzing in general, the different types of fuzzing
exhibit nuances in generating input and evaluating inputs based on results from monitoring the program.
We will next discuss these stages and their nuances.

\subsubsection{Pre-Fuzzing: Program Knowledge}
Prior to beginning the fuzzing process, a user incorporates program knowledge into the fuzzer.
At a minimum, a fuzzer needs program knowledge about how to \textit{instrument} or observe a program, including what constitutes an interesting program state, and about which input interfaces to explore.
For na\"ive fuzzers, this knowledge is sufficient to generate randomized input and send it to the indicated interfaces.
For non-na\"ive fuzzers, however, additional program knowledge is needed to effectively generate inputs.

\textbf{Mutation-based fuzzers} often require additional program knowledge in the form of an \textit{input corpus}, or a set of program inputs to feed to the indicated interfaces, in order to effectively generate inputs. 
Such program inputs tend to be examples of expected inputs, e.g., TLS handshake buffers for OpenSSL \cite{hanno} or memory dumps for Volatility \cite {Gaslight}.
Valid initial inputs guide the fuzzer to explore deep program states, i.e., states that are found after many branches in the program, by allowing the program to iterate on valid input, as opposed to random initial inputs that often do not pass initial program input checks.
Initial input corpora are generally created manually,
but large sets of inputs do pre-exist for some applications (e.g., all PDFs on the internet for Adobe Reader).

\textbf{Generation-based fuzzers} take in additional program knowledge in the form of input specifications, such as expected file formats or protocol descriptions.
Generation-based fuzzers typically achieve
better code coverage and deeper program state exploration than mutation-based fuzzers \cite{fell}.
These fuzzers are more difficult to set up because they require developing accurate program input specifications, which is often significantly more time-intensive than generating an input corpus. 
Generation-based fuzzers are limited to exploring the input space specified by these specifications.

\textbf{Evolutionary fuzzers}, like mutation-based fuzzers, take in additional program knowledge through an input corpus.

Typically, program knowledge is provided by the human user and is developed manually. 
In contrast, fuzzers automatically iterate over the next three stages of the fuzzing process. 
A fuzzer will generate new inputs, down-select and order inputs to send to the program, monitor the program for interesting program states, and repeat, generally until the user terminates the fuzzer. 
We explore these stages next. 

\subsubsection{Stage 1: Generate Inputs}
\label{sec:generate_inputs}
In the first stage of the fuzzing process, the fuzzer uses program knowledge to generate inputs to feed to the program through the identified interfaces. 
The goal of this stage is to create inputs that will explore new and interesting program states.  
Not all of the new inputs will be sent to the program; the next stage of the fuzzing process down-selects to the most relevant of the inputs generated here. 
However, the fuzzer aims here to generate the most relevant inputs.

\textbf{Mutation-based fuzzers} generate new inputs by manipulating previous ``interesting'' inputs.
In the first iteration of this stage, a fuzzer manipulates input provided in its input corpus.
Later, the fuzzer may adjust its collection of interesting inputs based on results from monitoring the program and evaluating input performance.

Mutation-based fuzzers mutate an interesting input by altering some portion of the input.
Two decisions are necessary for such mutation: 1) where to mutate (including the length of the mutation), and 2) what new value to use for the mutation \cite{rawat_17}.
Fuzzers use many techniques to make these decisions.
Common techniques include randomization (from bits to entire sections), specific bit flips, integer increments, and integer bound analysis and substitution. 

\textbf{Generation-based fuzzers} generate inputs by creating a new input according to the input specification.
Given a specification, a finite number of inputs corresponding to that specification exist \cite{takanen_08}. 
Because the input search space is finite, generation-based fuzzers can explore all possible specified inputs, which allows generation-based fuzzers to accurately measure how much of the input space they have explored.

\textbf{Evolutionary fuzzers}, again like mutation-based fuzzers, generate new inputs by manipulating previous interesting inputs.
Generally, evolutionary fuzzers either mutate one input or select two or more inputs and perform crossover, combining components of the selected inputs to make a new input; however, other randomization techniques may be used as well.
These fuzzers select inputs for randomization by evaluating input performance in previous stages.

Theoretically, mutation-based and evolutionary fuzzers could generate an infinite number of inputs.
The infinite input search space makes it difficult to estimate how much of the input space has been explored.
Additionally, as input lengths grow, mutation and crossover techniques require more manipulations to generate each new input and thus become more computationally challenging \cite{hthack}.
This can slow the fuzzing process considerably and severely limit the coverage obtained by the fuzzer in a set amount of time. 

\textbf{Symbolic Execution} is a static analysis technique that can help to generate new inputs that increase fuzzer coverage. 
Although symbolic execution is not a type of fuzzer, tools such as Driller~\cite{driller} and SAGE~\cite{godefroid_sage} have combined fuzzing and symbolic execution to improve input generation.
We discuss symbolic execution separately here because it can provide information for generating new inputs for mutation-based and evolutionary fuzzers. 
These new inputs can be added to the input corpus or otherwise fed to the program. 
We briefly describe how symbolic execution can help to generate new inputs to increase coverage. 

Symbolic execution analyzes a program to find \textit{constraints}, or limitations, on data values inside the program without making assumptions about input values fed to the program.\footnote{Actually, constraints can be added to input values as well, but we ignore that here for simplicity.}
Symbolic execution works by abstracting input data into \textit{symbolic} values, or data that might take on many \textit{concrete} values, and stepping through the program using these symbolic values.
When the program branches on a symbolic value, e.g., \texttt{if (x < 10)}, either side of the branch might be taken. 
In such cases, the symbolic execution engine makes two symbolic program states, each including a constraint met by the \textit{taken} side of the branch, and continues symbolically executing both of those states. 
In our example, the branch-taken state would include the constraint \texttt{x < 10} and the branch-not-taken state would include the constraint \texttt{x >= 10}; other than that, the two symbolic states would be identical. 

A symbolic state includes constraints representing the series of branch decisions necessary to reach that state, i.e., a path. 
A user can ask questions about the path used to reach that state. 
For example, what is a value of \texttt{x} that would reach this point?
To answer such a question, the symbolic state is \textit{modeled} as a set of constraints that encode, for each input variable, the range of valid values along the path. 
A constraint solver then \textit{solves} these constraints, either returning a valid concrete assignment to the input variables or proving that no such assignment exists (i.e., the constraints are unsatisfiable) \cite{raedt}. 
This assignment represents an input that would cause the program to reach the desired program state by making the same branch decisions as those in the original path. 

Fuzzers can use symbolic execution to create new inputs that explore new or interesting paths by asking the solver for solutions to newly discovered symbolic states. 
Unfortunately, symbolic execution is computationally costly, and the large number of possible program paths makes it infeasible to symbolically execute an entire program.\footnote{Paths or symbolic states are commonly exponential in the number of branches, and each iteration of a loop results in another branch.}
This high cost means that symbolic execution must be used cleverly; it is typically useful for finding paths that are difficult to find through random exploration. 
To mitigate this cost, fuzzers pair symbolic execution with their standard input generation techniques \cite{godefroid_sage}. 
The symbolic execution engine can limit paths by ``following'' an interesting input, e.g., only allowing new symbolic states to diverge from the input for a limited number of branches. 
However, the computational costs and path explosion remain significant hurdles. 
A large number of research efforts are attempting to address these
research challenges 
in symbolic execution \cite{driller, loop_sym_exe, path_exp_sym_exe}.  

Unfortunately, fuzzers still suffer from low coverage, even when pairing their specialized input generation techniques with symbolic execution.
Effective input generation remains a research challenge, and other clever applications of new techniques may help to intelligently generate inputs that increase coverage.
However, even these input generation techniques create many more inputs than fuzzers have computational time to execute. 
In the next stage, inputs are filtered to a smaller set to be fed into the program.

\subsubsection{Stage 2: Select Inputs}
In the second stage of the fuzzing process, the fuzzer selects and orders the inputs to send to the program. 
Recall, a fuzzer's goal is to exercise new program paths quickly. 
As B\"ohme et al. assert, however,  
most inputs exercise the same few program paths \cite{bohme_ccs_16, bohme_ccs_17}. 
To combat this, a fuzzer must use its input corpus effectively and minimize the computation spent to discover new program paths. 

\textit{Input test scheduling}, or \textit{seed selection}, can help combat this tendency of inputs to explore limited program paths. 
Input test scheduling ranks and selects inputs and input order, anticipating which new inputs are most likely to lead to new and disjoint interesting program states. 
For vulnerability assessment, test scheduling generally chooses inputs to maximize the number of bugs found \cite{woo}.
Input test scheduling is critical to effectively explore large or infinite input search spaces, but finding the right scheduling strategy for a particular program and fuzzer remains a research challenge.
Luckily, tools such as FuzzSim quickly compare input selection strategies using input performance information across many iterations of the process \cite{woo}.

As mentioned previously, as the fuzzer continues to iterate through the process, mutation-based and evolutionary fuzzers may collect many extremely large inputs. 
This artificially slows down the fuzzing process. 
To combat this, the fuzzer or user needs to perform either \textit{corpus minimization}, reducing the number of inputs, or \textit{input minimization}, reducing the size of each input. 
For example, \textit{corpus distillation}, introduced by Ormandy in ``Making Software Dumberer'' \cite{dumbsoftware}, does this by reducing a set of inputs to the minimal set of inputs that maintain the same coverage. 
The fuzzer of SEC Consult ignores specific portions of the input search space identified through manual analysis as not interesting in order to limit the number of manipulations performed on each input \cite{hthack}. 

Whether intelligently scheduling inputs, reducing the number of active inputs, or keeping inputs small, these techniques attempt to reduce the time needed for a fuzzer to find new interesting program states. 
However, effectively using the fuzzer's limited computational time remains a research challenge.
In the next stage, the fuzzer sends the selected inputs, monitors the program's resulting actions, and identifies interesting program states.

\subsubsection{Stage 3: Monitor Program}

In the third stage of the fuzzing process, the fuzzer feeds the program chosen inputs and monitors the program to identify interesting program states. 
``Interesting program states'' exhibit a specific program behavior in that state. 
In most cases, a crash represents the behavior of interest (i.e., the program fails unexpectedly). 
However, any behavior that is observable through program instrumentation might be used to identify interesting program states. 
For example, Valgrind can detect (observe) memory corruption even if the corruption does not cause a crash \cite{valgrind}. 
As another example, Heelan uses fuzzing to identify potential program memory allocators \cite{sheelan}. 
For vulnerability assessment, interesting behaviors are observable behaviors that relate to possible flaws or vulnerabilities. 

As mentioned, a fuzzer requires program knowledge about how to instrument the program and what constitutes an interesting program state. 
A human user often provides this information, but the definition of what an interesting program state \textit{should} be remains a research challenge.
In the vulnerability assessment case that is, what observable behaviors are most relevant to identifying flaws or vulnerabilities?

Once the fuzzer has identified an interesting program state, it needs to provide descriptions of that state back to the user for analysis. 
Program state descriptions differ wildly by fuzzer. 
For example, on a crash, one fuzzer might simply provide the input causing the crash, while another fuzzer might provide a full \textit{core dump}, a capture of program memory at the point of the failure.
After evaluation (the next stage), this information might guide further fuzzing iterations.

\subsubsection{Stage 4: Evaluate Inputs}
\label{sec:coverage_types}

In the fourth stage of the fuzzing process, the fuzzer evaluates how well inputs performed.
Many fuzzers use code coverage to measure the utility of an input: if the input causes execution of a new part of code (generally, a new basic block), that input has increased coverage and is rated highly.
The libFuzzer tool uses a similar metric, data coverage, which also rates an input highly if new data values occur at a previously explored comparison in the code.
Some fuzzers use bug discovery as a metric; inputs that cause a crash are rated highly.

\textbf{Evolutionary fuzzers} \textit{require} feedback about how well inputs performed \cite{fell}; that is, they must be able to evaluate inputs and rank them.
These fuzzers use input rank both in generating new inputs and in selecting and sending inputs.

\textbf{Mutation-based} and \textbf{generation-based} fuzzers do not generally require input performance feedback, but the same metrics can help to evaluate a fuzzer's performance overall.

Practically, coverage metrics are heuristics and do not provide a complete assessment of input performance.
Thus, effective and comparable metrics remain a research challenge.

\subsubsection{Post-Fuzzing: Interesting Program States}
\label{sec:postfuzzing}
Following the fuzzing process, the user analyzes interesting program states output by the fuzzer.
This process is often highly manual and benefits from research in related areas in software engineering. 
For each output state, the user analyzes the state to determine the \textit{root cause} of the interesting behavior in that state. 
Often the user manually observes as the program executes the associated input and hopes to identify the root cause.
The user then decides whether the root cause represents a new flaw or vulnerability. 

Often fuzzers output an overwhelming number of interesting program states.
The user must perform \textit{triage} to decide which interesting program states merit further investigation.
Unfortunately, output program states often share the same root cause.
Even worse, states that share the same root cause can appear drastically different.
For example, a memory corruption vulnerability can result in crashes in many different parts of the program, with wildly different memory images, because the effect of the vulnerability is not observed immediately. 
Some automated tools attempt to deduplicate fuzzer outputs \cite{sheelan} or root causes, but these are often imperfect, and triage and root cause analysis remain research challenges. 

\subsection{Comparing Fuzzers}
\label{sec:comparing_fuzzers}

Unfortunately, despite all the research into making fuzzers more effective and efficient, it is hard to determine whether one fuzzer is ``better'' than another.
Measuring the utility of a fuzzer is difficult for many reasons, from inability to explore the entire input search space which biases performance measures; to lack of ground truth, which makes validation a difficult and manual process; to randomness exploited in the fuzzing process, which introduces the need for statistical testing.
For example, in vulnerability analysis, fuzzers are often judged by number of unique vulnerabilities; however, when a new fuzzer finds new vulnerabilities in a set of real-world programs, this may simply show that the fuzzer's core algorithm is better suited to that set of programs than other algorithms \cite{rebert}. 
Results do not necessarily generalize well. 

To help judge fuzzer utility, Klees recently proposed a framework for fuzzer comparison \cite{hicks18}.
Klees calls for controlling comparisons by specifying: a baseline fuzzer, a benchmark suite of target programs, a performance metric, configuration parameters comparable to the baseline, and sufficient number of trials for statistical testing \cite{hicks18}.
Small data sets with known flaws can be used to compare and validate fuzzers in this way.
Unfortunately, however, these comparisons are computationally expensive, and, as mentioned, results do not necessarily generalize.
Practically, until thorough benchmarks for this framework have been developed, users will have to rely on their intuition to guide decisions about which fuzzer to use for a particular program, how to configure or adapt that fuzzer, and whether that fuzzer performs well enough.
Effective fuzzer comparison remains a significant research challenge.

\section{Applications of Machine Learning to Fuzzing}
\label{ML}
In this section, we explore research applying machine learning (ML) to the fuzzing process.
ML has been used to generate new inputs in the fuzzing process and, to a lesser extent, to improve post-fuzzing.  Unsupervised learning has seen the most successful applications to input generation, with fuzzing tools such as AFL \cite{lcamtuf} integrating genetic algorithms (GAs) into the input generation process.
There are also recent applications of both supervised and reinforcement learning (RL)  to input generation \cite{blum_arxiv_17, neuzz, becker, bottinger}.  Additionally, all three types of ML have been applied to symbolic execution \cite{bunz_gnn, wu_sat_ml, selsam, mairy}, primarily to reduce constraint equation solve times.
Both supervised and unsupervised learning have been applied to post-fuzzing processes primarily for crash triage and root cause categorization \cite{rebucket, harsh, long_patch}.  Interestingly, we know of no research in two areas: input minimization and corpus minimization.  These pieces of the fuzzing process tend not to be large bottlenecks, which may account for the lack of research.  

We begin by discussing ML research for input generation.  We then discuss applications of ML to post-fuzzing tasks.  Finally, we conclude by discussing fuzzing tasks that have not seen ML applications.  For each of these areas, we show how each type of ML offers unique methods for improvement as well as the difficulties of applying each type.

\subsection{Generate Inputs}
The most successful applications of ML to fuzzing occur in the generate inputs stage.  In this section we will discuss how the  various types of ML have been applied to input generation.  We will show that unsupervised learning, in the form of GAs and deep learning (DL), has had many successful applications to input generation.  We also discuss the applications of RL and supervised learning to various types of input generation, although this research tends to be more exploratory.

\subsubsection{Genetic Algorithms}
The most frequently used ML technique for input generation is GAs  \cite{fell, lcamtuf, liu, duchene,  demott}.  GAs, a type of unsupervised ML inspired by biological evolution, often provide the core input generation algorithms in evolutionary fuzzers.  There are 3 primary steps involved when using a GA: 1) generate a small base population of inputs, 2) perform transformations on inputs, and 3) measure the performance of the transformed inputs.  This process is repeated on the most performant inputs according to a chosen metric.  In the case of fuzzers, the base population of inputs consists of a set of seed program inputs.  The GA will mutate these seed inputs to explore the code space with the goal of discovering new paths in the code.  GAs have been successful for input generation due to this ability to build off previous successful inputs.

As mentioned in Section~\ref{sec:fuzzer_types} and further described in Section~\ref{sec:coverage_types}, evolutionary fuzzers use a fitness function to rank inputs for selection and mutation.  The choice of fitness function can have a tremendous impact on 1) the performance of the fuzzer, 2) the ability of the fuzzer to identify certain types of bugs, and 3) the tendency to get stuck in local minima.  Thus, particular care must be taken when choosing a fitness function.

While traditional code coverage is the most common metric used for the fitness function, more advanced heuristics such as the Dynamic Markov Model (DMM) heuristic have also been used \cite{sparks}.  In this work, the program control graph is represented as a Markov process, i.e. a stochastic process with transition probabilities for each edge of the control graph.  The DMM heuristic uses this Markov model and the transition probabilities to create a fitness function.  In contrast to code coverage, this setup allows for more precise control of the fuzzer, guiding it towards specific parts of the code that may contain a bug or flaw.

The choice of fitness function is a strong determinant of fuzzer behavior, and thus a promising area of research may be the development of new fitness functions which induce desired fuzzer behaviors.  
In particular, fitness functions which allow more fine-grained control of the fuzzer might be more beneficial for specific fuzzing goals.  Additionally, strategically combining or alternating between multiple metrics may also be useful in achieving fuzzing goals.

\subsubsection{Deep Learning and Neural Networks}

Deep learning (DL) and neural networks (NNs) have been applied for input generation, specifically to improve generation-based and mutation-based fuzzers.  Recurrent neural networks (RNN) \cite{rnn_survey} are the most common DL method applied to fuzzing.  In particular, Long-Short Term Memory (LSTM) \cite{rnn_survey} networks, a variant of the RNN, have been studied for input generation use.  

In one study, Godefroid et al. used LSTMs to create input grammar for PDF files in generation-based fuzzers \cite{godefroid_17}.  LSTMs have shown promise in many types of sequence generation tasks \cite{graves_2013}.  However, Godefroid et al. identified that the goals of fuzzing and LSTM input grammar learning often conflict with each other: LSTMs are biased towards producing well-formed inputs, while fuzzers aim to produce exemplar inputs that explore new program states.
To circumvent this paradox, a sampling strategy was used to pick inputs generated by the LSTM, ultimately creating an input grammar with a strong balance between well-formed and malformed inputs.  Experimental results demonstrated that using ML for input grammar generation is a promising technique for increasing code coverage. 

In a separate study, Rajpal et al. integrated DL into AFL to increase fuzzer coverage by selecting input bytes to mutate \cite{blum_arxiv_17}. An NN was used to generate a heat map indicating predicted likelihood of increasing code coverage when any particular byte is mutated.  Four different sequence learning architectures for generating the heatmap were compared: 1) the standard LSTM, 2) a bidirectional LSTM \cite{rnn_survey}, which processes input sequences both forwards and backwards, 3) Seq2Seq \cite{seq2seq}, which transforms a sequence to another sequence, and 4) a variant of Seq2Seq that uses an attention mechanism to focus on the most important parts of an input \cite{seq2seq_attn}. 
While each model increased code coverage in some situations, the standard LSTM model slightly outperformed the other models overall.  Experimentally the DL-augmented AFL outperformed standard AFL on the ELF, XML, and PDF formats, while standard AFL outperformed DL-augmented AFL on the PNG format.  

An alternative approach to generate inputs is to model program behavior, using the model to then select most promising inputs. 
The NEUZZ method used a shallow NN to model a program's behavior as a smooth continuous function \cite{neuzz}.  The NN was trained to predict the branching behavior of the program based on a seed input.  
Experimentally it was found that the gradients produced during the training process, specifically the larger gradients, were useful in identifying which input bytes control branching behavior.  NEUZZ used this information to guide fuzzer mutations, enabling it to experimentally outperform standard fuzzers including AFL and Angora on some programs.  

Chen et al. also modeled program behavior through the Angora fuzzing tool \cite{angora}.  Angora used a discrete function to represent the path from a program starting point to a specific branch constraint.  Gradient descent was then implemented over small discrete intervals within the function representation to find a set of inputs that satisfy the constraint and move the program through that particular branch. Overall, this method demonstrated the ability to quickly solve branch constraints and to experimentally outperform AFL and symbolic execution on some programs.

An additional alternative approach for input generation using DL was implemented by Cheng et al \cite{cheng_rnn}.  This approach used RNNs to predict new paths through a program.  These paths were then input into a Seq2Seq model, which generated new seed inputs for the fuzzer aimed to execute the predicted paths.  This preliminary study showed experimental that the generated corpus increased fuzzer code coverage for the PDF, PNG and TFF formats.    

DL and NN applications for input generation show promise, however there are still barriers to more common use.  First, DL models require significant computation time for training, so training likely cannot be practically integrated into the fuzzing process.  An alternative might be to train individual models for individual programs, but this is also likely impractical.  Many of these studies attempted to mitigate these issues.  For instance, Rajpal et al. circumvent training cost by training the NN on only a small corpus of inputs before fuzzing commences \cite{blum_arxiv_17}.  She et al., on the other hand, used a reduced form of model re-training throughout the fuzzing process by only retraining with the most useful data points \cite{neuzz}.  Both of these methods reduce the training time, but it is not clear what the general effect on fuzzing performance would be.  An alternative to reducing data size could be to develop methods to transfer previously trained models between programs.  Such methods might eliminate or reduce the amount of model training required for fuzzing previously unexplored programs, but it is not clear how these methods may effect performance.

A second barrier for DL application to fuzzing is performance consistency across file formats.  Many of these studies demonstrated that for some file formats, such as PDF, DL consistently improved fuzzing performance over prior baselines while other formats failed to compete with state-of-the-art techniques.  Future research for understanding the suitability of DL and NNs for specific file formats might be beneficial.

\subsubsection{Reinforcement Learning}

Two groups have applied reinforcement learning (RL) \cite{rl_survey}, \cite{deep_rl_survey} for input generation.  Becker et al. used the SARSA algorithm \cite{rl_survey} to fuzz the IPv6 protocol by mutating network packets sent to a host \cite{becker}.  The SARSA algorithm considers both the current state and action of the agent for determining optimal behavior.  Bottinger et al. used a deep Q-learning network \cite{q_learn} to learn a grammar describing inputs for generation-based fuzzers on the PDF format \cite{bottinger}.  The deep Q-learning network used a deep neural network to map states to actions.

These studies offer important insights to RL applications for input generation.
  First, they show that program representation is crucial for training a successful RL agent.  
Becker et al. used a finite state machine to represent the behavior of the IPv6 protocol where each state represented the current response of the host to a particular packet and transitions between states represent possible mutations to a packet.  
Bottinger et al. also used a type of finite state machine, the Markov decision process, for problem representation.  Markov Decision Processes are characterized by stochastic transitions between states.  In this study, the states represented a particular seed input for the fuzzer and the transitions represented probabilistic rewrite rules for the seed input at that state.  Thus, both Becker and Bottinger demonstrate the utility of using finite state machines as a problem representation for training an RL agent for input generation.  

Second, they offer insight into effectively defining a reward function for an agent, which is frequently the most challenging aspect of RL.  
Becker et al. created a multi-part reward function based on the following criteria: number of program functions called from a single input, presence of an error, and potential corruption or delay of the response message from the program \cite{becker}.  The presence of the error acted as the strongest signal to the agent that it had reached an interesting part of the code space.  The program response was used to guide the agent even in the absence of the error signal.  Thus, each part of the reward function played a unique role in guiding the agent and leaving out any of these criteria might have resulted in a less effective agent.
Bottinger et al. experimented with multiple distinct reward functions, one using code coverage, another using execution time, and a third combining code coverage and execution time. 
In both studies, the reward function influenced the fuzzer's exploration of the input space.  For instance, when Bottinger used execution time as a reward, the agent learned inputs that caused the program to terminate quickly.  The work by both Becker and Bottinger indicate a need to define reward functions carefully, taking into account the software program type, type of bugs being sought, fuzzing metrics available, and ultimate fuzzing goals. 

Standard fuzzers do not yet implement RL, and applications remain theoretical.  A key first step in this area is a more robust understanding of reward functions.  
Specifically, it is not yet clear how program dependent a reward function should be, or if there are reward functions that are generally effective. For instance, can the same reward function be used for every program or does each unique program require a unique reward function?  How does the file format (e.g., PDF and PNG) impact how a reward function should be defined?  

Another challenge for RL is understanding the transferability of an agent.  It is not currently clear if it is necessary to train a new RL agent for each unique program, and it is likely impractical to do so.
Therefore, studying and creating transferable agents is likely a requisite step for future research.  

\subsubsection{Machine Learning for Symbolic Execution}

Here, we briefly touch on different ML techniques currently being explored to improve symbolic execution.
As discussed in Section~\ref{sec:generate_inputs}, symbolic execution can help generate effective new inputs for fuzzers, but computational cost and path explosion remain significant hurdles. 
Several research efforts in ML explore the feasibility of improving constraint solving, which could support symbolic execution for fuzzer input generation. 
Unfortunately, current efforts in using ML do not rival state-of-the-art approaches using graph algorithms \cite{path_exp_sym_exe}; they are merely explorations in feasibility. 

Supervised learning has been used to solve constraint equations.  In one study, graph neural networks were used to identify features indicating whether constraint equations had valid solutions or not \cite{bunz_gnn}.  In another study, Wu used a combination of logistic regression and Monte Carlo methods to identify initial values that increased the probability of finding a valid solution to a constraint equation.  The Monte Carlo methods were used to identify initial promising values, while logistic regression was used to indicate the validity of the solution to the constraint equation using these chosen values.  Incorporating these new initial values led to decreased run-times for the Minisat solver \cite{wu_sat_ml}.  In another study, LSTMs (i.e., DL) were trained to solve constraint equations \cite{selsam}.  While the LSTMs were not able to beat state-of-the-art constraint solvers, they were able to solve constraint equations from domains they were not trained on, indicating an ability for the DL models to generalize.  Another method developed by Shiqi et al. used NNs as representations of constraint equations \cite{neuex}.  The solutions to these constraint equations were then discoverable via gradient descent.    In general, each of these studies offers unique ways of solving constraint equations.  While not competitive with current state-of-the-art, they nonetheless offer a strong starting point for using supervised learning to decrease the amount of computation time required to solve constraint equations.

Mairy et al. used RL to improve local neighborhood search methods \cite{mairy}.  Local neighborhood search methods iteratively find solutions to a constraint equation by finding solutions to various subsets of the constraint equation and combining the subsets to form the final solution.  In order to discover useful subsets, these local neighborhood search methods must intelligently explore the space of possible subsets.  The RL agent was guided to choose subsets that were more likely to lead to a valid solution, aiming to reduce the time needed to find such a solution.

One research effort used ML to reduce the size of the solution search space rather than to directly solve constraint equations.  Li et al. reformulated the typical collection of path constraints as an optimization problem and attempted to reduce the number of infeasible paths, i.e., paths that can never be reached due to conflicting constraints \cite{Li_2016}. 
They use the ML technique RACOS \cite{yu}, a technique for optimization that scales well to high dimensional problems, to solve the optimization problems; however, only very small programs of up to 335 lines were analyzed.

Initial explorations using ML to improve symbolic execution show promise; however, the practical utility of applying ML here remains to be seen. 
Current research is typically restricted to small problems that do not compete with state-of-the-art techniques. Additionally, this research has yet to be applied to the combined fuzzer and symbolic execution techniques described in Section~\ref{sec:generate_inputs}.  It is an open question if ML can aid in bettering these combined techniques.

\subsection{Post-Fuzzing: Interesting Program States}

As mentioned in Section~\ref{sec:postfuzzing}, users triage and then analyze interesting program states output by the fuzzer, often manually. 
Users triage program states by 1) evaluating for \textit{uniqueness} \cite{hercules}, 2) analyzing triaged states to determine \textit{reproducibility} and a root cause, and, 3) when fuzzing for vulnerability assessment, determining whether a root cause is exploitable.  ML has primarily been used to categorize crashes (triage) or to categorize bugs (root cause analysis), although Yan et al. used Bayesian methods and the \textit{!exploitable} tool to improve reliability for determining bug exploitability \cite{Yan2017ExploitMeterCF}. 

Dang et al. triaged crashes by grouping crashes with similar call stacks using agglomerative hierarchical clustering, an unsupervised learning technique that clusters data points with similar features \cite{rebucket}. 
They introduced their own similarity metric over call stacks, the position-dependent model, allowing them to use an unlabeled call stack data set for training. 
They tested their model over various Microsoft products and in many cases outperformed previous methods for crash similarity identification. 

There have been several attempts to use ML to categorize bugs in software.  Harsh et al. experimented with root cause categorization using a wide variety of supervised techniques including decision trees, support vector machines, and naive Bayes \cite{harsh}.  However, they note that applying supervised techniques can be challenging due to the lack of labeled data.  To combat the lack of labeled data, Harsh et al. also experiment with unsupervised and semi-supervised techniques.  Unfortunately, the techniques from this study are limited as the bug categories are extremely broad and system-specific.  

In another study, Long et al. used ML to both identify root causes and generate patches to repair the bug responsible \cite{long_patch}.  Their tool, Prophet, uses a parameterized log-linear probabilistic model which learns to identify important features that determine how a piece of code must be repaired.  One important aspect of this work is the interpretability of the Prophet model; the model parameters can be used to determine the importance of various features for generating a particular patch.  This interpretability is important as it helps the user understand why a particular patch was generated. 

Supervised ML techniques are used to identify root causes more frequently outside of the software domain \cite{sole}. 
In one instance decision trees and support vector machines were used for root cause analysis on industrial production systems \cite{Demetgul2013}. 
In another instance NN were used for fault localization within industrial tank systems \cite{sorsa_nn}. 
Support vector machines have also shown promise in speeding up fault localization on circuit boards \cite{ye_board}.
While none of this research was directly applied to software, it may be possible to extend the research to aid root cause analysis in the software domain.   

\subsubsection{Challenges}
 
ML is rarely applied to post-fuzzing tasks for two reasons: 1) ML results and algorithms are often difficult to interpret, and 2) appropriate training data sets are sparse.
For the first, most ML classification techniques return a prediction, not an explanation.
This makes it difficult for a user to determine whether a predicted label is correct and why that label was applied.
For example, in root cause analysis, the user would find it difficult to understand where the root cause supposedly manifested in the code, either to validate the label or to correct the root cause \cite{harsh}.
Further, ML algorithms often build up opaque rules that are difficult to map into domain knowledge. 

For the second, we have very few available labeled data sets, and it is not yet clear what constitutes a strong, generalizable benchmark data set.  
There are several questions without clear answers that must be addressed when constructing a data set for post-fuzzing tasks such as:
\begin{itemize}
\item Which of all the possible bugs should be included?
\item Which programming languages should be represented?
\item How should a bug and its root cause be encoded for an ML algorithm, especially given that root causes are nuanced and can vary from system to system?
\end{itemize}

\subsection{Seed selection}
There has been minimal research in \textit{seed selection} using ML.  Wang et al. conducted an extensive study using NNs to select inputs which were more likely to lead to vulnerabilities in the code being fuzzed \cite{neufuzz}.  While this study demonstrated promising results, challenges such as transferability of trained models to new programs still remain.  
As with input generation, model transferability remains a potential bottleneck for ML application to seed selection.  

Prior seed selection research points to another possibility for future ML research.  Seed selection must balance using current inputs with a known level of performance versus exploring new inputs with an unknown but potentially better performance  \cite{woo}.  RL algorithms are often applied successfully in these types of scenarios which require a balance of both exploration of new inputs and utilization of current inputs.  Thus, RL algorithms may be well suited for determining an optimal seed schedule.  Future research may include 1) creating reward functions for optimal seed selection, 2) determining transferability of offline RL agents between programs, and 3) integrating online RL agents within the fuzzing process.

\subsection{Input and Corpus Minimization}

To our knowledge, there has not been any research into \textit{input minimization} or \textit{corpus minimization} using ML.  We hypothesize two reasons for this lack of research.  First, neither input minimization nor corpus minimization is a large bottleneck.  The largest bottlenecks exist in the input generation and post-fuzzing process, thus, most research tends to concentrate on these areas.  Second, minimization of input or total corpus sizes does not naturally lend itself to ML techniques.  While input generation and, to a lesser extent, post-fuzzing are often able to be formulated as ML problems, minimization is often achieved successfully using heuristic methods \cite{dumbsoftware}.

\section{Conclusion}

In this survey, we discussed how machine learning (ML) has been applied to fuzzing. 
Because fuzzing problems lend themselves more naturally to unsupervised and reinforcement learning techniques, supervised learning is rarely used to support the fuzzing process.  ML is most often used to support the \textit{Generate Inputs} stage of the fuzzing process.  Unsupervised
methods currently offer the most benefit with tools such as AFL
making unsupervised algorithms part of their workflow.
In contrast, reinforcement learning and deep learning are being explored for possible improvements but are not yet commonly used. 
ML has also been applied to analyze \textit{Interesting Program States} during post-fuzzing, helping to triage crashes and support root cause analysis.
However, ML has not been applied to the \textit{Select Inputs} stage of the fuzzing process, possibly because this stage is not a major bottleneck.  Additionally ML has not been applied to evaluating reproducibility of crashes during post-fuzzing.  The lack of ML in certain portions of the fuzzing process is likely due to many factors including lack of training data, difficulty of transferring models between programs, and computationally intractable training times.

While ML has played an important role in the functionality of fuzzing systems, there remain many open research challenges.  Recently an influx of researchers have begun dedicating resources to address some of these challenges. 
Fuzzing will continue to play an important role in vulnerability assessment of programs in the future.  As research in this area grows, we expect to see the continued application of ML to address bottlenecks in the fuzzing process.

\begin{acks}
The authors would like to thank thank Danny Loffredo for insights into the practical application of fuzzing processes and Christopher Harrison for helpful discussions on symbolic execution.  The authors thank Michelle Leger for technical edits and discussions.  This work is funded in part by Sandia National Laboratories, a multi-mission laboratory managed and operated by National Technology and Engineering Solutions of Sandia, LLC., a wholly owned subsidiary of Honeywell International, Inc., for the U.S. Department of Energy's National Nuclear Security Administration under contract DE-NA-0003525.

\end{acks}

\bibliographystyle{ACM-Reference-Format}
\bibliography{ref}

\end{document}